\documentclass[reprint, amsmath, amssymb, aps, showkeys, nolongbibliography, numerical]{revtex4-2}

\usepackage{graphicx}
\usepackage[colorlinks=true, linkcolor=blue, allcolors=blue]{hyperref}
\usepackage[mathlines]{lineno}
\usepackage{xcolor}

\usepackage{xspace}
\newcommand\AFIVE{{AMP5}\xspace}
\newcommand\ATEN{{AMP10}\xspace}

\begin{document}

\title{Conformal invariance in water-wave turbulence}

\author{M. Noseda}
\email{mnoseda@df.uba.ar}
\affiliation{Universidad de Buenos Aires, Facultad de Ciencias Exactas y Naturales, Departamento de Física, Ciudad Universitaria, 1428 Buenos Aires, Argentina}
\affiliation{CONICET - Universidad de Buenos Aires, Instituto de Física Interdisciplinaria y Aplicada (INFINA), Ciudad Universitaria, 1428 Buenos Aires, Argentina}

\author{P.~J. Cobelli}
\email{cobelli@df.uba.ar}
\affiliation{Universidad de Buenos Aires, Facultad de Ciencias Exactas y Naturales, Departamento de Física, Ciudad Universitaria, 1428 Buenos Aires, Argentina}
\affiliation{CONICET - Universidad de Buenos Aires, Instituto de Física Interdisciplinaria y Aplicada (INFINA), Ciudad Universitaria, 1428 Buenos Aires, Argentina}

\date{Last edited: \today}

\begin{abstract}
  We experimentally investigate the statistics of zero-height isolines in gravity wave turbulence as physical candidates for conformal invariant curves.
  We present direct evidence that they can be described by the family of conformal invariant curves called stochastic Schramm-L\"owner evolution (or SLE$_{\kappa}$), with diffusivity $\kappa = 2.88(8)$. A higher nonlinearity in the height fields is shown to destroy this symmetry, though scale invariance is retained.
\end{abstract}


\maketitle

Symmetries in physics underpin fundamental laws and reveal profound connections between seemingly unrelated phenomena. Scale invariance, for instance, establishes links between disparate processes, ranging from phase transitions, galaxy clustering, and crack propagation to avalanches in granular media, and social networking. In certain systems exhibiting critical behavior, conformal invariance emerges as a broader symmetry in which the properties remain unaltered under conformal transformations, i.e., geometrical transformations that preserve angles and allow for non-uniform rescaling of distances.

Conformal symmetry plays a central role in understanding critical phenomena, scaling behaviors, and emergent properties across
classical and quantum field theory domains —from particle physics to condensed matter, turbulence, and string theory \cite{itzykson1998conformal,henkel2013confcrit}. In 2D, conformal invariance completely determines the scaling exponents and correlation functions of a wide class of theories at criticality \cite{cardy_non-equilibrium_2008}, enabling a thorough classification of universality classes of critical phenomena.

In recent years, the discovery of Schramm-L\"owner Evolution (SLE) has provided physicists with a powerful and versatile framework for the statistical characterization of conformal invariance in physical systems \cite{schramm_scaling_2000,cardy_sle_2005,bauer2006,henkel2012conformal}. SLE is a class of non-self-intersecting random planar curves that can be mapped into a one-dimensional Brownian walk with diffusivity $\kappa$, and thus have conformal invariant statistics. Moreover, SLE$_{\kappa}$ gives rise to a natural classification (by the value of $\kappa$) of conformal curves in the plane, such as the boundaries of clusters in 2D critical processes described by conformal field theories.

Studies based on the SLE formalism have revealed surprising connections between diverse branches of physics. Pioneering research showed, for example, that the zero-vorticity lines in the inverse cascade of 2D Navier-Stokes turbulence are conformal invariant and statistically equivalent to the boundaries of percolation clusters \cite{bernard_conformal_2006,puggioni_conformal_2020}, whereas temperature isolines in surface quasi-geostrophic turbulence were found to belong to the same universality class as domain walls in the O(2) spin model \cite{bernard_inverse_2007}. Conformal invariance was also reported for   nodal lines of random wave functions \cite{keating2006,bogomolny_sle_2007}, domain walls of spin glasses \cite{amoruso_conformal_2006,bernard_possible_2007}, rocky shorelines \cite{boffetta_how_2008}, isoheight lines on growing solid surfaces \cite{saberi_conformal_2008,saberi_conformal_2008-1}, avalanche frontiers in sandpile models \cite{saberi_direct_2009}, a class of active scalar turbulence \cite{FalkovichMusacchio}, isovorticity lines in 3D rotating turbulence \cite{thalabard_conformal_2011}, watersheds dividing drainage  basins \cite{daryaei_watersheds_2012}, graphene sheets \cite{giordanelli_conformal_2016}, and critical percolation clusters \cite{pose_shortest_2014,DeCastro2018}.

Turbulence is defined as the state of a physical system with many interacting degrees of freedom deviated far from equilibrium. The main fundamental inquiry in turbulence concerns the degree of universality and the symmetries of the turbulent state; particularly which symmetries remain broken and which ones are restored in the developed stage \cite{Falkovich_symmetries}. In this context, the aforementioned SLE-based studies demonstrated the emergence of conformal invariance in various dual cascade turbulent systems, all characterized by the prevalence of vortices \cite{bernard_conformal_2006,puggioni_conformal_2020,FalkovichMusacchio,thalabard_conformal_2011}.
Since wave turbulence systems are known to exhibit properties similar to those of their vortex counterparts (such as scale invariance, universality, and dual cascades), the question naturally arises: Can conformal invariance be observed in wave turbulence scenarios, i.e., in turbulent systems dominated by the coupling between random nonlinear waves? This subject is particularly important for its potential implications, as wave turbulence is present in a vast range of physical processes and scales; from the ocean state to nonlinear optics, from quantum fluids to gravitational waves
\cite{zakharov1992kolmogorov,nazarenko2011wave,shrira2013advances,galtier_2022}.

In this Letter we address this question experimentally for the case of water-wave turbulence, a dual cascade system \cite{zakharov1992kolmogorov} theorized to present conformal invariance \cite{cardy_non-equilibrium_2008}.
Our results show that the zero-height isolines of weak water-wave turbulence are, within experimental uncertainty, compatible with stochastic Schramm-L\"owner evolutions, with statistics remarkably close to those of domain walls in the critical 2D Ising model \cite{chelkak_convergence_2014}.

In our experiments, water-wave turbulence is generated by the horizontal motion of two piston-type wavemaker paddles (150 mm in width, 30 mm immersed) within a $(1000 \times 800)$~mm$^{2}$ open-top PMMA transparent tank filled with the working fluid up to a rest height of 50~mm. The wavemakers are independently controlled by two linear servomotors, with a positional repeatability of $\pm\,0.05$~mm and 44~N peak force. Their motion is subjected to a random forcing (both in amplitude and phase) with a white frequency spectrum bandpass-filtered in the 0--4~Hz range, with adjustable maximum amplitude. This forcing scheme is known to produce cascades of water-wave turbulence in laboratory experiments
\cite{Falcon2007b,Herbert2010,cobelli_different_2011,Falcon2022a}.
In this study, we investigated the effects of two peak forcing amplitudes: 5 mm (\AFIVE) and 10 mm (\ATEN); the latter representing a four-fold increase in driving power relative to the former.

We employ diffusive light photography (DLP) to obtain 3D measurements of the topography of the wave field. This optical technique \cite{Wright1996a, wright_imaging_1997} allows us to measure the local free-surface height $h(x,y,t)$ in a region of $(241.5 \times 241.5)$~mm$^{2}$. The surface dynamics is recorded by a high-speed camera with a resolution of $(1024 \times 1024)$~px$^2$, at 60~Hz  and 1/3000 s shutter speed (for experimental details, see Supplemental Material, Sec. I  \cite{[{See Supplemental Material at }][{ for more information }] SUPMAT}, which includes Refs. \cite{ishimaru1978wave,berhanu_space-time-resolved_2013,Henry2000,Xia2012,Francois2014,Francois2017}). 
Each experimental realization comprises 1700 consecutive frames, covering a duration of 28.3~s. Throughout our experimental campaign, each dataset consists in five independent realizations conducted under identical experimental conditions. 

\begin{figure}[t!]
\centering
\includegraphics[scale=0.9]{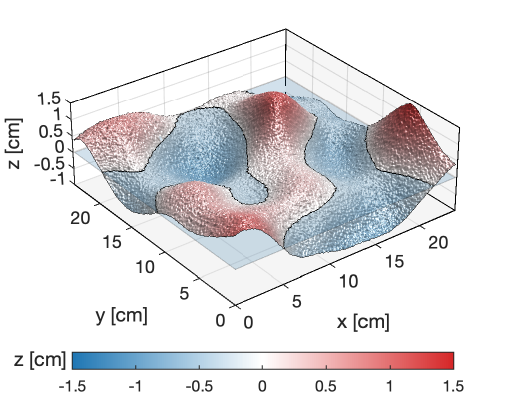}
\caption{Snapshot of the reconstructed height field $h(x,y)$. As a visual reference, the position of the free surface at rest (corresponding to $z=0$) is represented by a translucent blue plane. Zero-height isocontours of $h(x,y)$ are depicted, onto the free surface, by solid black lines.}
\label{fig:free-surface-snapshot}
\end{figure}

Figure~\ref{fig:free-surface-snapshot} shows an example of free-surface reconstruction obtained by the DLP technique. Individual isocontours of zero height, depicted in the Figure as black curves over the $z=0$ plane, are derived from the reconstructed fields using Matplotlib's {\it contour} function \cite{Hunter2007}. This function employs a marching squares algorithm and linear interpolation to define the isolines as a series of discrete points in the $(x, y)$ plane. By construction, each contour line is oriented so that positive height sites are consistently positioned to its left.

From the complete set of zero-height isolines, candidate SLE traces are identified as follows. Firstly, for each free-surface reconstruction, a horizontal line representing a real $x$-axis in the complex plane is drawn across the height field. Any contiguous portion of a zero-height isoline that lies between two successive intersections with the real axis is considered a candidate trace. This process is repeated for all four possible orientations of each height map, corresponding to incremental rotations in steps of $\pi / 2$ radians.

To avoid using time-correlated paths, we only select candidate traces corresponding to height fields separated by $n$ frames, using $n = 10$ and only keeping traces that exceed lengths of $l = 500$ points (larger values of both $n$ and $l$ yield similar results). Following this procedure, we collected a total of 1154 candidate traces for \AFIVE, with an average of 1908 points per trace. Similarly, \ATEN yielded 1337 putative traces, each consisting of an average of 1637 points.

\begin{figure}[t!]
\centering
\includegraphics[scale=0.9,trim={0 5 0 0}, clip]{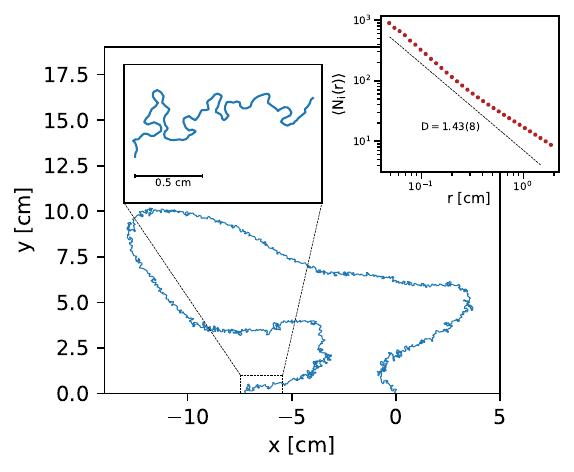}
\caption{Example candidate trace and fractal dimension estimation for the \AFIVE dataset. A close-up view on the last points of the trace is presented in the left inset, highlighting the complex structures present at smaller scales. The right panel shows, for the \AFIVE dataset, the power-law scaling of the (mean) number of sticks required for travelling the traces as a function of their length (circles).}
\label{fig:fractal_dimension}
\end{figure}

Figure~\ref{fig:fractal_dimension} shows an example  candidate trace belonging to the \AFIVE dataset. An initial observation reveals a rich and intricate structure, hinting at the existence of underlying complexity. Moreover, a close-up view on an arbitrarily chosen part of it (depicted in the upper-left inset) evidences the persistence of such structures across different scales, suggesting the presence of self-similarity or fractal-like characteristics.

In order to investigate the scaling behavior of the candidate traces we
estimate their fractal dimension by means of Richardson's yardstick method
\cite{Mandelbrot1967, *Mandelbrot1983}. In this method, a yardstick of length $r$ is considered and the length of the candidate trace is defined by the number of straight yardsticks $N(r)$ required to go from one extreme to the other by jumping from one point on the curve to the next at a distance $r$. The fractal (or capacity) dimension $D_{0}$ is defined as the limit $D_{0} = - \lim_{r \rightarrow {0^{+}}} \log N(r) / \log(r)$.
Prior to its application to candidate traces, we validated our implementation of the yardstick algorithm with synthetic traces \cite{kennedy_numerical_2009,foster_asymptotic_2022} (for details, see Supplemental Material  \cite{[{See Supplemental Material at }][{ for more information, which includes Refs. XXX}] SUPMAT}).

For each candidate trace (indexed by $i$), $N_{i}(r)$ was computed across a scale range spanning the dataset minimum to maximum point separation. The ensemble average at a fixed scale, denoted as $\langle N_{i} \rangle(r)$, is presented in the top-right panel of Fig.~\ref{fig:fractal_dimension} for the \AFIVE dataset. A power-law relation is evident at small $r$, spanning a decade of scales within the available range.
The best fit to the data (depicted as a dotted line), results in $D_{0} = 1.43 \pm 0.08$. A similar analysis for the \ATEN dataset yields $D_{0} = 1.35 \pm 0.07$ (not shown). These results indicate that candidate traces exhibit statistical self-similarity across the resolved scales, providing strong support for scale invariance in our datasets; a necessary (though not sufficient) condition for conformal invariance.

To probe the conformal invariance of these curves, we consider a self-avoiding curve $\gamma(t)$, parameterized by the time $t$, which begins at the origin and extends to $\infty$. At any time $t$, the upper half-plane $\mathbb{H}$ minus the curve up to that point, can be mapped back onto $\mathbb{H}$ by an analytic function $g_{t}(z)$, made unique on imposing the asymptotic condition
$g_{t}(z) \sim z + {2t}/{z} + O(1 / z^{2}) $ for $z \rightarrow \infty$.  The growing tip of the curve is therefore mapped to a point $\xi(t)$ on the real axis. The curve $\gamma(t)$ and the conformal map $g_{t}(z)$ are completely parametrized by the driving function $\xi(t)$, and satisfy the chordal L\"owner equation \cite{Lowner1923}:%
\begin{equation}
    \frac{d g_t}{dt} = \frac{2}{g_t(z) - \xi(t)}.
    \label{eq:loewner}
\end{equation}%
In the case of random curves, Eq.~\eqref{eq:loewner} is called Schramm--L\"owner evolution (SLE), and the driving function is a random variable.
Schramm demonstrated that the statistics of random curves are conformal invariant if and only if the driving function is a Brownian walk \cite{schramm_percolation_2001}, i.e., a random function with uncorrelated increments and satisfying $\sigma_{\xi}^{2} = \langle \left[ \xi(t) - \langle \xi(t) \rangle \right]^2 \rangle = \kappa t$. Here, $\kappa$ is the diffusivity which allows for the classification of the conformal invariant random curves into universality classes denoted by SLE$_{k}$. 

Therefore, in order to test the conformal invariance of the zero-height isolines, it is necessary to assess whether their driving functions are statistically a gaussian process, and specifically if it corresponds to Brownian motion (termed direct test). If this were the case, the diffusivity $\kappa$ should also be determined.

The first step consists in extracting the driving functions of our candidate SLE traces by inversion of the L\"owner equation.
As Eq.~\eqref{eq:loewner} is valid for chordal curves, the traces given by the set of points $\{z_{0}, z_{1}, \ldots, z_{N}\}$ are first translated to the origin and then mapped using a M\"obius transformation of the form $\varphi(z) = z_{N} z / (z_{N}-z)$, to ensure that the curves start at the origin and end at infinity. We verified that this procedure does not modify our results greatly.

Assuming that the ensemble of zero-height isolines is statistically equivalent to SLE curves, we derived the associated driving functions ensemble using the zipper algorithm with vertical slits \cite{marshall_convergence_2007,Kennedy2007} (chosen amongst the many available numerical algorithms \cite{Kennedy2008}). This algorithm is based on the solution of Eq.~\eqref{eq:loewner} for a simple trace made of an infinitesimal vertical line segment, given by: $g_{\xi_{t}, \delta t}(z) = \xi_t + \sqrt{(z-\xi_t)^2+4 \, \delta t}$. This conformal transformation maps the upper half-plane, excluding the vertical slit joining the points $\xi_{t}$ and $\xi_{t} + 2i \sqrt{\delta t}$, back into the upper half-plane. The process begins with $z_{0} = (0,0)$, and all subsequent points except the first one are mapped
by $g_{\xi_{t}, \delta t}(z)$, with $\delta t = \frac{1}{4} \operatorname{Im}(z_{1})^{2}$ and $\xi_{t} = \operatorname{Re}(z_{1})$. The resulting set of points is renumbered, resulting in a curve one element shorter than the original. By repeating this
procedure on the remaining points, the algorithm zips the entire trace onto the real axis by the composition of conformal maps. After mapping all points to the real axis, this results in a sequence of discrete L\"owner times $t_{j}$ and driving values $\xi_{t_{j}}$ that approximate the true driving functions. Lastly, the final L\"owner time for all traces is renormalized to 1, using the scaling property of SLE \cite{Lawler_inbook}.

\begin{figure}[t!]
\centering
    \includegraphics[scale=0.9,trim={0 10 0 0}, clip]{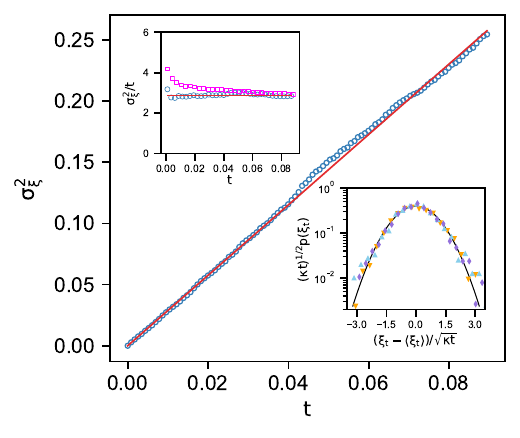}
    \caption{Statistics of the ensemble of driving functions. Main frame: linear behavior of $\sigma_{\xi}^{2}(t)$ for traces in the \AFIVE dataset, with a slope of $\kappa = 2.88(8)$. Upper-left inset: diffusivity $\kappa$, in circles and squares for the \AFIVE and \ATEN datasets, respectively. Lower-right inset: PDF of the rescaled driving functions for curves in the \AFIVE dataset, at three different times: $t = 0.0015$ (inverted triangles), $0.045$ (triangles), and $0.09$ (diamonds); the solid line represents a standard Gaussian distribution.}
    \label{fig:direct-test}
\end{figure}

We begin by considering the \AFIVE dataset. An ensemble average over the driving functions allows us to observe a linear growth of the variance $\sigma_{\xi}^{2} = \langle \left[ \xi_t - \langle \xi_{t} \rangle \right]^{2} \rangle$ with L\"owner time, as shown in Fig.~\ref{fig:direct-test} (circles). To quantify this observation, we perform a fit of the form $\sigma_{\xi}^{2}(t) = \kappa\, t^{\alpha}$ within that time range, represented by a solid line in the Figure. Notably, our analysis confirms the initial observation, yielding a value of $\alpha = (1.00 \pm 0.01)$, providing strong support for the linear scaling hypothesis. Moreover, we find the best estimate for the diffusivity to be $\kappa = (2.88 \pm 0.08)$.
The compensated value of the variance, i.e. $\sigma^{2}_{\xi}(t) /t \sim \kappa(t)$, is shown in the upper-left inset of Fig.~\ref{fig:direct-test} in circles, along with the best estimate from the fit, represented by a solid line. A plateau that extends throughout the range of L\"owner times is clearly observed.

In contrast, repeating this analysis with the \ATEN dataset yields qualitatively different results. In particular, for this dataset the variance of the driving functions does not scale linearly with L\"owner time.
This is evidenced by the lack of a plateau in the compensated variance for this dataset, represented by squares in the top-left inset of Fig.~\ref{fig:direct-test}. %

We associate this breakdown of conformal invariance to an increase in the level of nonlinearity in the interaction between waves. This becomes apparent when examining the typical root mean square wave steepness, $\sigma_s$, defined as the standard deviation of the magnitude of the gradient of $h$ \cite{berhanu_space-time-resolved_2013}. Remarkably, the \ATEN dataset exhibits a $\sigma_s$ value of 0.21, nearly double that of the \AFIVE dataset, which is  characterized by $\sigma_s = 0.11$. The heightened nonlinearity manifests also in the distribution of wave height values, with \ATEN displaying a greater departure from Gaussianity compared to \AFIVE (see Supplemental Material, Sec. II \cite{[{See Supplemental Material at }][{ for more information }] SUPMAT}).

We proceed by testing whether $\xi_{t}$ is a Gaussian process. 
We conducted Kolmogorov-Smirnov tests on the standardized driving
$\left[\xi_{t} - \langle \xi_{t} \rangle \right] / (\kappa t)^{2}$, separately for each (discrete) L\"owner time available within the range where linearity was established. 
The results indicate that, at a significance level of 0.05, the null hypothesis of Gaussianity is not rejected in any of the individual tests. Upon combining the $p$-values
from these multiple tests using Fisher's method, the combined $p$-value
of 0.78 does not provide sufficient evidence to reject the null hypothesis. This suggests that the driving process
$\xi_t$ appears consistent with a Gaussian distribution. Correspondingly, the PDF of the standardized driving collapses onto a standard Gaussian distribution, as illustrated in the lower-right inset of Fig.~\ref{fig:direct-test} for three L\"owner times. 

For conformal invariant traces, the fractal dimension and the diffusivity of the associated SLE$_{k}$ process are related by $D_{0} = \min \lbrace 2, 1 + \kappa / 8 \rbrace$ \cite{beffara_dimension_2008}. As we have independently determined both $D_{0}$ and $\kappa$ for the zero-height isolines, this theoretical prediction serves as a consistency check for our findings. Remarkably, the fractal dimension derived from that expression using $\kappa = 2.88(8)$ yields a value of $D_{0}^{\smash{(\kappa)}} = (1.36 \pm 0.01)$, in agreement with the one previously obtained.

\begin{figure}[t!]
\centering
\includegraphics[scale=0.9,trim={0 7 0 0}, clip]{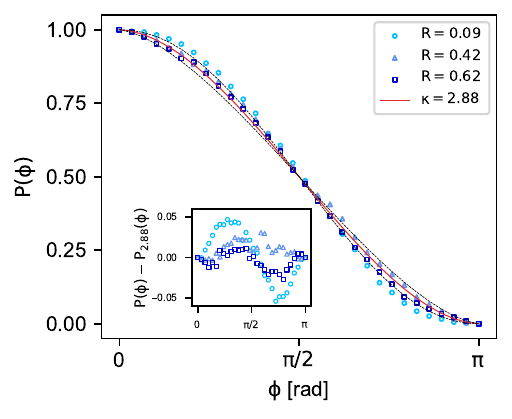}
    \caption{Left passage probability, $P(\phi)$, for the zero-height isolines of wave turbulence, for different radii within the variance linear regime. The solid line corresponds to the theoretical prediction for $\kappa = 2.88$, the diffusivity value obtained by the direct test; dashed lines represent the edges of $10 \%$ envelopes around that value. The corresponding residuals $P(\phi)-P_{2.88}(\phi)$, are shown in the inset.}
    \label{fig:left-passage-probability}
\end{figure}

We finally consider the left passage probability (LPP). This property quantifies the probability $P_{\kappa}(\phi)$ of a chordal SLE$_{\kappa}$ trace passing to the left of a given point $R \, e^{{i \phi}}$ in $\mathbb{H}$, where $R$ and $\phi$ are the distance and the angle between the origin and the point. Due to scale invariance, $P_\kappa(\phi)$ is independent of $R$, and SLE theory \cite{schramm_percolation_2001} predicts that%
\begin{equation}
    P_\kappa(\phi) = \frac{1}{2} + \frac{\Gamma\left(\frac{4}{\kappa}\right)}{\sqrt{\pi}\: \Gamma\left(\frac{8-\kappa}{2\kappa}\right)} \: _2F_1 \left(\frac{1}{2}, \frac{4}{\kappa}, \frac{3}{2}, -\cot^2 \phi \right) \cot(\phi),
\label{eq:left_passage_probability_prediction}
\end{equation}%
where $_2F_1$ and $\Gamma$ represent the Gauss hypergeometric and Gamma functions, respectively.
For this test, the traces are regenerated from their individual driving functions with the ensemble mean  $\langle \xi_{t} \rangle$ substracted.
The LPP is calculated as the (normalized) number of curves passing to the left of a point in the upper half-plane.
Figure~\ref{fig:left-passage-probability} presents the results for  $P(\phi)$ for traces in the \AFIVE dataset. To assess its independence from radial distance, we systematically computed the LPP for different radii, three of which are shown in the Figure. Those specific radii, chosen to illustrate the typical behavior observed for the LPP, correspond on average to L\"owner times located at the initial, intermediate and final stages of the linear scaling regime shown in Fig.~\ref{fig:direct-test}.
The analytical prediction $P_{2.88}(\phi)$ from Eq.\eqref{eq:left_passage_probability_prediction} with $\kappa = 2.88$ is depicted by a solid line, while dashed lines represent $\pm 10\%$ of this value. The inset displays the residuals $P(\phi) - P_{2.88}(\phi)$ for the radii considered in the main Figure. Our results for the LPP exhibit good agreement with the theoretical prediction for $\kappa = 2.88$.

In summary, we presented an experimental study that shows compatibility between SLEs and zero-height isolines of water-wave turbulence. This argument is supported by four mutually independent standardized tests performed on our data, namely: linearity of the variance with L\"owner time (direct SLE test), gaussianity assesment through Kolmogorov-Smirnov testing, fractal dimension of the traces, and left passage probability. Within the \AFIVE dataset, all tests are in agreement with SLE predictions, and lead to numerically consistent results for the diffusivity. In particular, our results yield $\kappa = 2.88(8)$, a value close to that of domain walls in the critical 2D Ising model, with $\kappa = 3$ \cite{chelkak_convergence_2014}. Although this cannot be interpreted as proof, the extension of theoretical prediction \cite{cardy_non-equilibrium_2008} with our experimental findings constitutes a robust indication for conformal invariance in the weakly coupled limit.  In contrast, the AMP10 dataset, characterized by higher forcing, larger deviations from Gaussianity in the distribution of wave height, and heightened nonlinearity (as evidenced by its wave steepness), does not demonstrate compatibility with the SLE family (yet retains scale invariance).

The observation that the zero-height isolines are conformal invariant places the study of water-wave turbulence within the theory of critical phenomena, and suggests that at least part of its statistics could be described in terms of a conformal field theory. Finally, our results naturally bring forward the question as to whether other systems exhibiting wave turbulence may also present conformal invariance as a restored symmetry, and if so, to what universality classes would they belong. This can open a new approach to the general study of the broad class of nonlinearly interacting systems described by wave turbulence.

\begin{acknowledgements}
The authors acknowledge support from Grants Proyectos de Investigación Científica y Tecnológica No. 2018-4298, and UBACyT No. 20020170100508. Moreover, the authors are grateful to Pablo Mininni for fruitful discussions.
\end{acknowledgements}

\end{document}


\title{Supplemental material:\\ Conformal invariance in water-wave turbulence}
\author{M. Noseda}
\author{P. J. Cobelli}%
\affiliation{Departamento de Física, Facultad de Ciencias Exactas y Naturales, UBA and INFINA, CONICET, Ciudad Universitaria, 1428 Buenos Aires, Argentina}%

\date{\today}
            
\maketitle

\section{Experimental setup and measuring technique}

As the main article describes, for measuring the wave height of our system we worked with diffusing light photography (DLP). This technique 
relies on the transport mean free path
of the light being larger than the wave height of the field under study, but smaller than the fluid depth. Under these conditions, light does not propagate in rays but instead scatters, diffusing through the bulk \cite{ishimaru1978wave}, locally linking the fluid depth to the intensity of the transmitted light. As a consequence, the light intensity transmitted through the liquid, $I(x,y)$, is related to the depth of the fluid with respect to the rest height, $h(x,y)$, through the relation: $ I(x,y) = I_0(x,y) \exp \left[- A \, h(x,y) \right]$, where $I_{0}$ represents the intensity distribution of light transmitted through the fluid layer at rest, and
$A$ is the absorption coefficient.

These conditions are usually met experimentally by adding a low concentration of a diffusing agent to water, such as plastic microspheres \cite{Wright1996a,wright_imaging_1997}, lipidic emulsions \cite{berhanu_space-time-resolved_2013} or milk \cite{Henry2000,Xia2012,Francois2014,Francois2017}. In this study we work with a 6\,\% volumetric concentration of homogenised milk (3\,\% fat content) in distilled water. 
Such low dilution does not significantly modify the fluid viscosity and density from pure water values. The surface tension of the solution was measured statically by a Krüss K8 Du Noüy ring tensiometer, providing a value of $(45 \pm 2)$~mN/m. This leads to a capillary length of 2.15~mm (i.e., 20\,\% lower than that of pure water). The transition between gravity and capillary waves is expected to occur for a critical wavenumber $k_{c} = \sqrt{\rho g / \gamma}$, which corresponds to a gravity--capillary crossover frequency of 15.22~Hz.

\begin{figure}[t!]
\centering
\includegraphics[width=0.45\textwidth]{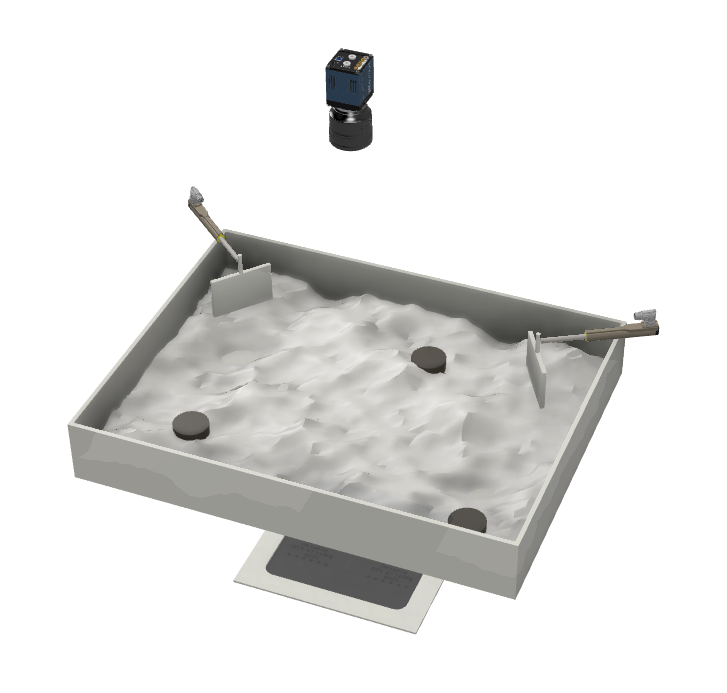}
    \caption{Schematic diagram of the experimental setup, showing the LED panel (at the bottom), the wavetank with the linear servomotors forcing the paddles, and the camera collecting the light that traverses the working fluid.}
    \label{setup}
\end{figure}

The fluid is illuminated from below by a centrally positioned light-emitting diode panel ($400 \times 400$~mm$^{2}$) placed underneath the transparent bottom of the tank. A Photron 1024PCI fast camera, located 1 m above the free-surface, records the transmitted light in a region of area $(241.5 \times 241.5)$~mm$^{2}$ at the center of the illuminated region. The free-surface dynamics is sampled at 60~Hz, using a shutter speed of 1/3000~s. The horizontal spatial resolution, corresponding to $(1024 \times 1024)$~px$^{2}$, is 0.24~mm. To mitigate camera photon shot noise, captured images undergo spatial filtering through a Gaussian kernel with unit standard deviation. 

The absorption coefficient of the working fluid is calibrated before each experiment, by gradually increasing the liquid depth and measuring the change in transmitted light intensity when the liquid surface is flat. The dynamic range of our images (10 bits) and their contrast allows us to resolve a 30~$\mu$m change in the fluid elevation.

Figure \ref{setup} shows an schematic diagram of the experimental setup used in this work. As the wavetank is rectangular, several obstacles in the form of surface-piercing cylinders are located randomly (sufficiently far from the observation region) in order to avoid the presence of cavity modes.

\section{Wave height and steepness distributions}

In this Section we present the PDF distributions of height and wave steepness for the \AFIVE and \ATEN datasets.

\begin{figure}[t!]
\centering
    \includegraphics[width=0.45\textwidth]{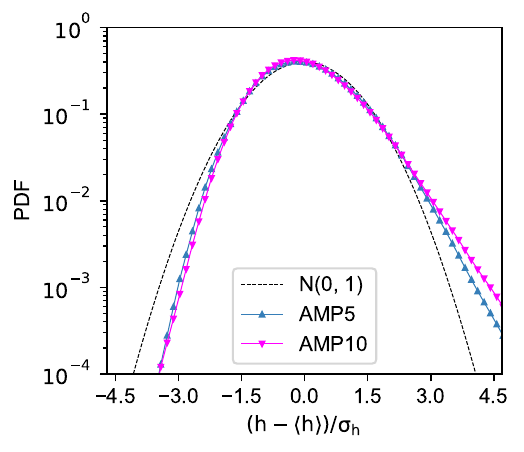}
    \caption{Probability density functions (PDF) of the renormalised height, $(h-\langle h \rangle) / \sigma_h$, for \AFIVE and \ATEN datasets. A standard Gaussian distribution is shown in dashed lines, for reference.}
    \label{histograms}
\end{figure}

The wave height distributions are shown in Figure \ref{histograms}, where a standard Gaussian distribution is also shown for comparison purposes. The distributions are seen to depart from Gaussianity as the amplitude of the forcing is increased. For comparison, the values of the skewness and excess kurtosis for \AFIVE are 0.37 and 0.30 respectively, with a typical wave height of $\sigma_h = 4.9$~mm, whereas for \ATEN those values reach 0.66 and 1.01 respectively, with a typical wave height of $\sigma_h = 8.0$~mm.

For random wavefields such as those in our study, the typical RMS wave steepness is defined by \cite{berhanu_space-time-resolved_2013} 
\begin{equation*}
\sigma_s^r = \bigg\langle \sqrt{ \frac{1}{\mathcal{S}} \int_\mathcal{S} || \nabla h ||^2 \: dx \: dy  - \left( \frac{1}{\mathcal{S}} \int_\mathcal{S} || \nabla h || \: dx \: dy \right)^2  } \bigg\rangle,
\label{eq:wave_steepness*}
\end{equation*}

\noindent where $|| \nabla h || = \sqrt{ (\partial_x h)^2 + (\partial_y h)^2}$, and the $r$ superscript stands for {\it random}. As this definition corresponds to calculating the standard deviation of $||\nabla h(x,y,t)||$, computed over all points in space for every time-instant available, we also studied the distribution of the magnitude of the gradient of height.

For the evaluation of the numerical derivatives $(\partial_x h, \partial_y h)$, we first applied a 2D Gaussian filter to smooth the height fields and reduce the effects associated with noise in numerical differentiation. This smoothing filter amounts to the 2D convolution of the height maps $h(x,y,t)$, composed of $1024 \times 1024$ points, with a Gaussian kernel of characteristic size $\sigma = 5$. We verified that the  results described below remain insensitive to this choice beyond that value (e.g., $\sigma = 10$ yields consistent quantitative results), provided the kernel size is not excessively large. The gradient is then computed from the smoothed fields by means of second-order accurate central differences in the interior points and first-order accurate one-side differences at the boundaries.

\begin{figure}[t!]
\centering
    \includegraphics[width=0.45\textwidth]{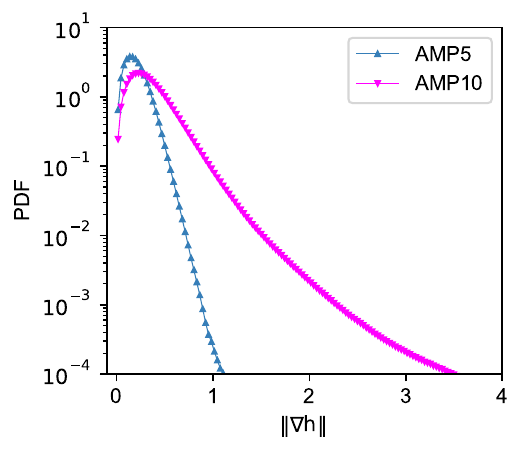}
    \caption{Probability density functions (PDF) of the magnitud of the surface gradient for both datasets. The typical (rms) wave steepness of \AFIVE is $\sigma_s = 0.11$ whereas for \ATEN its value reaches $\sigma_s = 0.21$.}
    \label{PDFws}
\end{figure}

The probability density functions for $||\nabla h||$ for both datasets are shown in Fig.~\ref{PDFws}. The PDFs show qualitatively different behaviors, with longer tails present in the AMP10 dataset, reaching values as high as 3 for the magnitude of the gradient (with associated probabilities of order $\sim 10^{-4}$).
In contrast, the PDF of $||\nabla h||$ for AMP5 presents a more abrupt fall-off, with values below 1.

The associated typical (rms) wave steepness calculated from the aforementioned expression is $\sigma_s^r = 0.11$ for AMP5, whereas for AMP10 it is approximately twice that value: $\sigma_s^r= 0.21$.
These values are also close to the classical (monochromatic) estimate of wave steepness given by $\sigma_s^m = \sigma_h k_c$. Here,  $k_c = 20.7$~rad/m, is the wavenumber obtained from the linear gravity-capillary dispersion relation for $f_c = 2$~Hz, the central frequency of the forcing range in our experiments. For such evaluation we obtain a wave steepness of 0.10 and 0.17 for AMP5 and AMP10, respectively.

In summary, both estimations of wave steepness (monochromatic and random) lead to  values above 0.1, indicating that the level of nonlinearity in our experiments is not small. In particular, the distribution of $||\nabla h||$ for the AMP10 dataset shows the presence of steeper slopes, and is characterized by a typical (rms) wave steepness that doubles that of the AMP5 dataset. Table~\ref{tab:wave_height_steepness} summarizes the main quantitative results for the distributions of wave height and steepness for both datasets.

\begin{table}[t!]
    \centering
  \label{Table 1}
    \begin{tabular}{lccccccccc}
         & & $\sigma_h$ [mm] & & $\tilde{\mu}_3^h$ && $ \tilde{\mu}_4^h-3$ &&& $\sigma_s$ \\ \hline
        \AFIVE  & & {$4.9$} & & {$0.37$} && {$0.30$} && &{$0.11$} \\
        \ATEN  & & {$8.0$} & & {$0.66$} & &{$1.01$} && &{$0.21$}
    \end{tabular}
    \caption{Height and wave steepness distributions' summary statistics for both datasets \AFIVE and \ATEN. From left to right, the columns report the wave height standard deviation $\sigma_h$, skewness and excess kurtosis (with tildes denoting standardized moments); as well as the wave steepness $\sigma_s$. }
    \label{tab:wave_height_steepness}
\end{table}

\section{SLE testing}

For testing compatibility between our isolines and SLEs we developed a Python code that calculates the best fit for the value of the diffusivity according to three theoretical predictions: the fractal dimension, the scaling of the variance of the driving functions (or direct test) and the left passage probability. In the following subsections we describe how this code was validated with synthetic traces.

\subsection{Construction of the synthetic datasets for testing}

For checking the performance of these three tests we employed two different SLE simulated ensembles. The first one was generated using Kennedy's algorithm \cite{kennedy_numerical_2009}, and the second one was built by means of Foster code \cite{foster_asymptotic_2022}. From each, we generated two synthetic datasets with $\kappa = 1$ and with $\kappa = 4$. For future reference, the traces generated using Kennedy's code are termed K1 and K4 respectively, and the ones built from Foster's algorithm are called F1 and F4, respectively. Both simulations from Kennedy correspond to an ensemble of 2000 traces with exactly 2000 points each. The simulations by Foster correspond to 2000 traces but their lengths (in points) varies because of the way the algorithm constructs them; the traces from F1 have an average number of points of 2117 while the ones from F4 have an average of 7516 points.

\subsection{Fractal dimension}

For estimating the fractal dimension we implemented a yardstick algorithm \cite{Mandelbrot1983}, which is useful when working with planar curves.
We tested this algorithm using a known fractal curve such as the Koch curve with seven iterations, which yielded a fractal dimension of $D_0 = 1.27$, close to its theoretical fractal dimension given by $\textrm{log}(4)/\textrm{log}(3) \simeq 1.26$.

We also applied it to our synthetic traces (described precedently) and obtained the results shown in Table \ref{fd}. The predicted value for the diffusivity according to Beffara's prediction \cite{beffara_dimension_2008} is also shown in Table \ref{fd}, showing good agreement with the theoretical value.
\begin{table}[h]
    \centering
  \label{Table 1}
    \begin{tabular}{ccccc}
         & \textbf{{K1}} & \textbf{{F1}} & \textbf{{K4}} & \textbf{{F4}} \\ \hline
        $D_0$  & {$1.17$} & {$1.18$} & {$1.46$} & {$1.55$} \\
        \textbf{$\kappa$}  & {$1.37$} & {$1.46$} & {$3.74$} & {$4.42$}
    \end{tabular}
    \caption{Values for the fractal dimension $D_{0}$ calculated with our implementation of the yardstick method, and corresponding diffusivity $\kappa$ for the traces in the synthetic datasets.}
    \label{fd}
\end{table}

\subsection{Direct SLE test}

The direct test consists in verifying if the driving functions of the traces are compatible with Brownian motions. For SLE curves in the plane, the variance of the ensemble of the driving functions follows $\sigma^2_{\xi} = \kappa t$. We obtained the driving functions using the zipper algorithm with vertical slits (see main text for details and references therein), and then interpolated these functions in a discrete L\"owner time grid. At each grid point we calculated the variance of the ensemble of driving functions and using a least squares fit we obtained the slope which corresponds to the best estimate for the diffusivity. In Table \ref{dt} we show these results for the simulations, also in good agreement with their expected value.

\begin{table}[h]
    \centering
  \label{Table 1}
    \begin{tabular}{ccccc}
         & \textbf{{K1}} & \textbf{{F1}} & \textbf{{K4}} & \textbf{{F4}} \\ \hline
        \textbf{$\kappa$}  & {$0.97$} & {$1.03$} & {$3.96$} & {$4.04$}
    \end{tabular}
    \caption{Values of the diffusivity obtained for the synthetic datasets with our implementation of the direct SLE test.}
    \label{dt}
\end{table}

\subsection{Left passage probability}

The left passage probability tests the probability that the traces pass to the left of a point $R e^{i\phi}$,  where $R$ and $\phi$ are the distance and the angle between the point and the origin, respectively. The function that the probability follows according to the SLE theory is given by Eq. (2) in the main text.
We found the best value for $\kappa$ by minimising the integral of the absolute value of the residuals. Those values are shown in Table \ref{lpp}.

\begin{table}[h]
    \centering
  \label{Table 1}
    \begin{tabular}{ccccc}
        & \textbf{{K1}} & \textbf{{F1}} & \textbf{{K4}} & \textbf{{F4}} \\ \hline
        \textbf{$\kappa$}  & {$1.03$} & {$1.02$} & {$3.75$} & {$3.96$}
    \end{tabular}
    \caption{Values of the diffusivity for the synthetic datasets according to the left passage probability.}
    \label{lpp}
\end{table}